\documentclass[conference]{IEEEtran}
\usepackage{graphicx}
\usepackage{algorithm2e}
\usepackage{algorithmic}
\usepackage{amsmath}
\usepackage{amsfonts}
\usepackage{amssymb}
\begin{document}
\title{Simple and Efficient Secret Sharing Schemes for Sharing Data and Image}
\author{\IEEEauthorblockN{Binu V.P}
\IEEEauthorblockA{Department of Computer Applications\\
Cochin University\\
Kochi, India\\
binuvp@gmail.com}
\and
\IEEEauthorblockN{Sreekumar A}
\IEEEauthorblockA{Department of Computer Applications \\
Cochin University\\
Kochi, India\\
sreekumar@cusat.ac.in}}
\maketitle
\begin{abstract}
Secret sharing is a new alternative for outsourcing data in a secure way.It avoids the need for time consuming encryption decryption process and also the complexity involved in key management.The data must also be protected from untrusted cloud service providers.Secret sharing based solution provides  secure information dispersal by making shares of the original data and distribute them among different servers.Data from the threshold number of servers can be used to reconstruct the original data.It is often impractical to distribute data among large number of servers.We have to achieve a trade off between security  and efficiency.An optimal choice is to use a $(2,3)$ or $(2,4)$ threshold secret sharing scheme, where the data are distributed as shares among three or four servers and shares from any two can be used to construct the original data.This provides both security,reliability and efficiency.We propose some efficient and easy to implement secret sharing schemes in this regard based on number theory and bitwise XOR.These schemes are also suitable for secure sharing of images.Secret image sharing  based on Shamir's schemes are lossy and involves complicated Lagrange interpolation.So the proposed scheme can also be effectively utilized for lossless sharing of secret images.
\end{abstract}
\section{Introduction}

The secret sharing schemes are originally proposed by Shamir \cite{shamir1979} and Blackley \cite{blakley1979} in 1979.
The motivation was to safeguard cryptographic keys.Their solution was to store the secret keys at several locations as shares and when authorized number of users collaborate together, they can retrieve the secret.The schemes are $(t,n)$ threshold schemes where any $t$ number of users can collaborate to recover the secret out of $n$ users.This provides both security,reliability and convenience.Shamir's scheme is simple and easy to implement and is based on polynomial interpolation.Blackley's scheme has a different approach and is based on hyperplane geometry.But it is difficult to implement.Secret sharing schemes have found numerous applications in designing several cryptographic protocols .\@ Threshold cryptography \cite{desmedt1992shared}, access control \cite{naor1998access}, secure multi-party computation \cite{ben1988completeness} \cite{chaum1988multiparty} \cite{cramer2000general}, attribute based encryption \cite{goyal2006attribute} \cite{bethencourt2007ciphertext}, generalized oblivious transfer \cite{tassa2011generalized}   \cite{shankar2008alternative}, visual cryptography  \cite{naor1995visual} $etc.,$ are some of the important areas where secret sharing schemes are used.In this paper we suggest efficient secret sharing schemes for the reliable and secure distributed storage of data on untrusted servers.

Shamir's scheme is based on polynomial interpolation over a finite field. It uses the fact that we can construct a polynomial of degree $t-1$ only if  $t$ data points are given.The scheme is based on polynomial interpolation.Given $t$ points in the 2-dimensional plane $(x_i,y_i),\ldots,(x_t,y_t)$, with distinct $x_i$'s, there is one and only one polynomial $P(x)$ of degree $t-1$ such that $P(x_i
)=y_i$ for all $i$.In order to share the secret $S$ , pick a random $t-1$ degree polynomial $P(x)=a_0+a_lx+ \ldots +a_{t-1}x^{t-1}$ with $a_0=S$,and evaluate shares as $S_1'= P(1),S_2'=P(2),\ldots S_i' = P(i)\ldots S_n' = P(n)$.Any subset of $t$ of these shares $S_i'$ together with their identifying indices, we can find the coefficients of $P(x)$ by interpolation, and then evaluate $S=P(0)$.The knowledge of just $t-1$ of these values,  does not suffice in order to calculate $S$.Efficient $O(n{log^2n})$ algorithms exist for the evaluation and interpolation of polynomials.

A secret sharing scheme is called perfect if less than $t$ shares give no information about the secret.It is known that for a perfect secret sharing scheme $H(S_i) \geq H(S)$. If $H(S_i) = H(S)$ then the secret sharing scheme is called ideal.
Shamir's scheme is perfect and ideal.Blackley's scheme is not perfect.

Confidentiality,reliability and efficiency are the major concerns in secure storage of data.The idea of secret sharing for the information dispersal is suggested by Krawczyk et al \cite{Kraw1994ssmdshrt} in 1994.He proposed a computationally secure secret sharing scheme for the distributed storage using Rabin's \cite{rabin1989ida} information dispersal algorithm and Shamir's secret sharing scheme.However the data is encrypted using a symmetric key encryption and the share of the key is distributed along with the data shares.The share size is less than the secret in this case compromising the information theoretic security.Abhishek Parak et al \cite{Abhi2010spaceff} in 2010 proposed a space efficient secret sharing scheme for
the implicit data security.They incorporated $k-1$ secrets in $n$ shares and any $k$ shares can be used to reconstruct the original secret.A recursive construction using Shamir's scheme is applied in which computational over head is more.Recursive methods of secret sharing is also mentioned in \cite{gan2002recrsv} , \cite{pra2010treerec}.Computational secret sharing schemes are proposed for the space efficiency in  \cite{philp1995css},\cite{rog2007robcss},\cite{vin2003powrcss}.

Secret sharing based solution provides information theoretical security on confidentiality with out encryption and hence avoid the complexities associated with encryption and key management.It also provides the guarantee on availability of data.Perfect secret sharing needs large amount of computational overhead.We propose specially designed secret sharing schemes using XOR and number theoretic technique to reduce the computation overhead.Unanimous consent schemes are easy to implement using XOR.But the implementation of a general $(t,n)$ threshold scheme is difficult.Wang et al \cite{wang2007ssboln} proposed a scheme based on boolean operation which is used for secret image sharing in 2007.Kurihara et al \cite{kuri2009xor3n},\cite{kuri2009xor} proposed a $93,n)$ and a generalized $(t,n)$ secret sharing scheme based on simple XOR operations.Efficient and ideal threshold scheme based on XOR is proposed by Lv et al \cite{Lv2010effidlxor} in 2010.
Secret sharing using number theoretic schemes are also developed based on Chinese reminder theorem \cite{asmuth1983},\cite{mignotte1983},\cite{Iftene2007crtss}.They are not widely used because of the computational complexity.The proposed scheme make use of simple number theoretic concept and the Euclid's algorithm.

\section{Proposed Secret Sharing Schemes}
The proposed system suggests a method of storing and retrieving private data in a secure and effective manner. The private data include personal information, sensitive information or unique identification etc. The data storage may be a private information storage using cloud database.We propose number theoretic and XOR based scheme for efficient implementation of secret sharing scheme.It can be used for secure storage and retrieval.Since it does not involve any encryption, the PKI needed for key management can be avoided.Section 2.1 contains the detailed description of the secret sharing algorithm using number theoretic concept.Section 2.2 explains the XOR based schemes.The algorithms mentioned below are designed to share one byte of data at a time.The scheme can be used to share both textual data and images.

\subsection{Schemes Based on Number Theory}
In this section the proposed secret sharing schemes which are based on number theoretic concepts and are explained in detail.Two threshold secret sharing schemes of order $(2,3)$ and $(2,4)$  are proposed.The Algorithm \ref{alg:twobythree} is the $(2,3)$ secret sharing phase and the retrieval algorithms depend on which shares are used for the reconstruction and are given in Algorithms \ref{alg:s1s2}, \ref{alg:s1s3},\ref{alg:s2s3}.A $(2,4)$ secret sharing scheme is mentioned in Algorithm \ref{alg:twobyfour}.The secret revealing algorithms corresponds to different combination of shares are given in
Algorithms \ref{alg:s1s2-24},\ref{alg:s1s3-24},\ref{alg:s1s4-24},\ref{alg:s2s4-24},\ref{alg:s3s4-24}.The algorithms use simple number theory concept.In order to find the inverse of a number extended Euclid's algorithm can be used.The share generation can be done with a complexity of $O(n)$ and the secret revealing can be done with a complexity of $O(nlog n)$, where $n$ is the number of bytes to share. Table lookup can be used for faster performance.

\RestyleAlgo{boxruled}
\SetAlgoNoLine
\begin{algorithm}[!ht]
\label {alg:twobythree}
\KwData{Input file S to share.}
\KwResult{Three Shares S1,S2,S3 of same size as the original file.}
\BlankLine
Choose a field $Z_p$ where $p=257$. \\
\While{not at end of the input file}{
$s$=read\_byte(S)  // read a byte or pixel \\
\If{$s==0$}{
$s=256$
}
$a=s^\frac{p-1}{3}$ //find cube root of s \\
$r$=random(257) // random number between 0-256 \\
$s1=r \times a \; \mbox{mod}\; p$ // s1 is the share1 pixel\\ 
\If{$s1==256$}{
$s1=0$
}
$s2=r^2 \times a \; \mbox{mod}\; p$ // s2 is the share2 pixel\\ 
\If{$s2==256$}{
$s2=0$
}
$s3=r^4 \times a \; \mbox{mod}\; p$ // s3 is the share3 pixel\\ 
\If{$s3==256$}{
$s3=0$
}
}
\caption{(2,3) Secret Sharing: Number Theory}
\end{algorithm}

\begin{algorithm}[!ht]
\label {alg:s1s2}
\KwData{Shares S1 and S2}
\KwResult{The original secret file S which is shared}
\BlankLine
Choose a field $Z_p$ where $p=257$. \\
\While{not at end of the input files}{
$s1$=read\_byte(S1)  // read a byte or pixel from S1\\
$s2$=read\_byte(S2)  // read a byte or pixel from S2\\
\If{$s1==0$}{
$s1=256$
}
\If{$s2==0$}{
$s2=256$
}
$a=s1^2 \times s2^{-1}\; \mbox{mod} \;p$ \\
$s=a ^3\; \mbox{mod}\; p$; // s is the secret data byte or pixel
\If{$s==256$}{
$s=0$
}
}
\caption{(2,3) Secret Revealing:Number Theory S1S2}
\end{algorithm}

\begin{algorithm}[!ht]
\label {alg:s1s3}
\KwData{Shares S1 and S3}
\KwResult{The original secret file S which is shared}
\BlankLine
Choose a field $Z_p$ where $p=257$. \\
\While{not at end of the input files}{
$s1$=read\_byte(S1)  // read a byte or pixel from S1\\
$s3$=read\_byte(S3)  // read a byte or pixel from S2\\
\If{$s1==0$}{
$s1=256$
}
\If{$s3==0$}{
$s3=256$
}
$s=s1^4 \times s3^{-1}\; \mbox{mod} \;p$ // s is the secret data byte or pixel \\
  
\If{$s==256$}{
$s=0$
}
}
\caption{(2,3) Secret Revealing:Number Theory S1S3}
\end{algorithm}

\begin{algorithm}[!ht]
\label {alg:s2s3}
\KwData{Shares S2 and S3}
\KwResult{The original secret file S which is shared}
\BlankLine
Choose a field $Z_p$ where $p=257$. \\
\While{not at end of the input files}{
$s2$=read\_byte(S2)  // read a byte or pixel from S1\\
$s3$=read\_byte(S3)  // read a byte or pixel from S2\\
\If{$s2==0$}{
$s2=256$
}
\If{$s3==0$}{
$s3=256$
}
$a=s2^2\times s3^{-1}\; \mbox{mod} \;p$ \\
$s=a ^3\; \mbox{mod}\; p$; // s is the secret data byte or pixel
 
\If{$s==256$}{
$s=0$
}
}
\caption{(2,3) Secret Revealing:Number Theory S2S3}
\end{algorithm}

\begin{algorithm}[!ht]
\label {alg:twobyfour}
\KwData{Input file S to share.}
\KwResult{Four Shares S1,S2,S3,S4 of same size as the original file.}
\BlankLine
Choose a field $Z_p$ where $p=257$. \\
\While{not at end of the input file}{
$s$=read\_byte(S)  // read a byte or pixel \\
\If{$s==0$}{
$s=256$
}
$r$=random(257) // random number between 0-256 \\
$s1=r$ // s1 is the share1 pixel \\ 
\If{$s1==256$}{
$s1=0$
}
$s2=r \times s \; \mbox{mod}\; p$ // s2 is the share2 pixel\\ 
\If{$s2==256$}{
$s2=0$
}
$s3=r^2 \times s \; \mbox{mod}\; p$ // s3 is the share3 pixel\\ 
\If{$s3==256$}{
$s3=0$
}
$s4=r^3 \times s \; \mbox{mod}\; p$ //s4 is the share4 pixel \\
\If{$s4==256$}{
$s4=0$
}
}
\caption{(2,4) Secret Sharing:Number Theory}
\end{algorithm}
\begin{algorithm}[!ht]
\label {alg:s1s2-24}
\KwData{Shares S1 and S2}
\KwResult{The original secret file S which is shared}
\BlankLine
Choose a field $Z_p$ where $p=257$. \\
\While{not at end of the input files}{
$s1$=read\_byte(S1)  // read a byte or pixel from S1\\
$s2$=read\_byte(S2)  // read a byte or pixel from S2\\
\If{$s1==0$}{
$s1=256$
}
\If{$s2==0$}{
$s2=256$
}
$s=s1\times s2^{-1}\; \mbox{mod} \;p$ \\
\If{$s==256$}{
$s=0$
}
}
\caption{(2,4) Secret Revealing:Number Theory S1S2}
\end{algorithm}
\begin{algorithm}[!ht]
\label {alg:s1s3-24}
\KwData{Shares S1 and S3}
\KwResult{The original secret file S which is shared}
\BlankLine
Choose a field $Z_p$ where $p=257$. \\
\While{not at end of the input files}{
$s1$=read\_byte(S1)  // read a byte or pixel from S1\\
$s3$=read\_byte(S3)  // read a byte or pixel from S3\\
\If{$s1==0$}{
$s1=256$
}
\If{$s3==0$}{
$s3=256$
}
$s=(s1^2)^{-1}\times s3\; \mbox{mod} \;p$ \\
\If{$s==256$}{
$s=0$
}
}
\caption{(2,4) Secret Revealing:Number Theory S1S3}
\end{algorithm}
\begin{algorithm}[!ht]
\label {alg:s1s4-24}
\KwData{Shares S1 and S4}
\KwResult{The original secret file S which is shared}
\BlankLine
Choose a field $Z_p$ where $p=257$. \\
\While{not at end of the input files}{
$s1$=read\_byte(S1)  // read a byte or pixel from S1\\
$s4$=read\_byte(S4)  // read a byte or pixel from S4\\
\If{$s1==0$}{
$s1=256$
}
\If{$s4==0$}{
$s4=256$
}
$s=(s1^3)^{-1}\times s4\; \mbox{mod} \;p$ \\
\If{$s==256$}{
$s=0$
}
}
\caption{(2,4) Secret Revealing:Number Theory S1S4}
\end{algorithm}

\begin{algorithm}[!ht]
\label {alg:s2s3-24}
\KwData{Shares S2 and S3}
\KwResult{The original secret file S which is shared}
\BlankLine
Choose a field $Z_p$ where $p=257$. \\
\While{not at end of the input files}{
$s2$=read\_byte(S2)  // read a byte or pixel from S2\\
$s3$=read\_byte(S3)  // read a byte or pixel from S4\\
\If{$s2==0$}{
$s2=256$
}
\If{$s3==0$}{
$s3=256$
}
$s=s2^2\times s3^{-1}\; \mbox{mod} \;p)$ \\
\If{$s==256$}{
$s=0$
}
}
\caption{(2,3) Secret Revealing:Number Theory S2S3}
\end{algorithm}

\begin{algorithm}[!ht]
\label {alg:s2s4-24}
\KwData{Shares S2 and S4}
\KwResult{The original secret file S which is shared}
\BlankLine
Choose a field $Z_p$ where $p=257$. \\
\While{not at end of the input files}{
$s2$=read\_byte(S2)  // read a byte or pixel from S2\\
$s4$=read\_byte(S4)  // read a byte or pixel from S4\\
\If{$s2==0$}{
$s2=256$
}
\If{$s4==0$}{
$s4=256$
}
$s=sqrt(s2^3\times s4^{-1}\; \mbox{mod} \;p)$ \\
\If{$s==256$}{
$s=0$
}
}
\caption{(2,4) Secret Revealing:Number Theory S2S4}
\end{algorithm}
\begin{algorithm}[!ht]
\label {alg:s3s4-24}
\KwData{Shares S3 and S4}
\KwResult{The original secret file S which is shared}
\BlankLine
Choose a field $Z_p$ where $p=257$. \\
\While{not at end of the input files}{
$s3$=read\_byte(S3)  // read a byte or pixel from S2\\
$s4$=read\_byte(S4)  // read a byte or pixel from S4\\
\If{$s3==0$}{
$s3=256$
}
\If{$s4==0$}{
$s4=256$
}
$s=s3^3\times (s4^2)^{-1}\; \mbox{mod} \;p$ \\
\If{$s==256$}{
$s=0$
}
}
\caption{(2,4) Secret Revealing:Number Theory S3S4}
\end{algorithm}

\subsection{Schemes based on XOR}
An $(n,n)$ scheme using XOR can easily be setup by creating
$n-1$ random shares of same size as the secret and the $n$th share as the XOR of these $n-1$ shares and the secret $k$.The secret can be revealed by simply XOR ing all the shares.In this we propose two scheme.An ideal $(2,3)$ scheme where the size of the share is same as that of the secret is mentioned in Algorithm \ref{alg:xoridlshr} and a non ideal scheme which is also not perfect is mentioned in Algorithm \ref{alg:secshrxor}
In this the size of the share is reduced to half.The scheme can be used when the storage become a constraint.The secret sharing and revealing can be done in time $O(n)$, where $n$ is the number of bytes to share.The secret reconstruction corresponds to different combination of shares in the non ideal scheme are mentioned in Algorithms \ref{alg:xor12},\ref{alg:xor13},\ref{alg:xor23} and in the ideal schemes are mentioned in Algorithms \ref{alg:xoridlrec12},\ref{alg:xoridlrec13},\ref{alg:xoridlrec23}.

\begin{algorithm}[!ht]
\label {alg:secshrxor}
\KwData{Secret file S to share.}
\KwResult{Three shares S1,S2 and S3 of half the size of S.}
\BlankLine
\While{not at end of the input files}{
$s$=read\_byte(S)  // read a byte or pixel from S\\
$bs$=binary($s$)  // bs is the binary representation of s\\
// odd bits of bs taken as share1 data nibble s1
$s1$=odd\_bits$(bs)$ \\
// even bits of bs taken as share2 data nibble s2
$s2$=even\_bits$(bs)$ \\
//share3 nibble is formed by xoring s1 and s2
$s3=s1 \oplus s2$ 
}
\caption{(2,3) XOR secret sharing-non ideal}
\end{algorithm} 
\begin{algorithm}[!ht]
\label {alg:xor12}
\KwData{Share S1 and S2}
\KwResult{The original secret file S which is shared.}
\BlankLine
\While{not at end of the input files}{
$s1$=read\_byte(S1)  // read a byte or pixel from S1\\
$s2$=read\_byte(S2)  // read a byte or pixel from S2\\
$s=\mbox{intermix}(s1,s2)$ // intermix the bits of s1 and s2 to construct the secret byte
}
\caption{(2,3) XOR secret revealing S1S2-non ideal}
\end{algorithm}
 
\begin{algorithm}[!ht]
\label {alg:xor13}
 \KwData{Share S1 and S3}
 \KwResult{The original secret file S which is shared.}
 \BlankLine
 \While{not at end of the input files}{
 $s1$=read\_byte(S1)  // read a byte or pixel from S1\\
 $s3$=read\_byte(S3)  // read a byte or pixel from S3\\
 $s2=s1 \oplus s3$ \\
 // intermix the bits of s1 and s2 to construct the secret byte\\
 $s=\mbox{intermix}(s1,s2)$ // intermix the bits of s1 and s2 to construct the secret byte
 }
 \caption{(2,3) XOR secret revealing S1S3-non ideal}
 \end{algorithm}
 
 \begin{algorithm}[!ht]
 \label {alg:xor23}
  \KwData{Share S2 and S3}
  \KwResult{The original secret file S which is shared.}
  \BlankLine
  \While{not at end of the input files}{
  $s2$=read\_byte(S2)  // read a byte or pixel from S2\\
  $s3$=read\_byte(S3)  // read a byte or pixel from S3\\
  $s1=s2 \oplus s3$ \\
  $s=\mbox{intermix}(s1,s2)$ // intermix the bits of s1 and s2 to construct the secret byte
  }
  \caption{(2,3) XOR secret revealing S2S3-non ideal}
  \end{algorithm}
  \clearpage
  \newpage
 
  \begin{algorithm}[!ht]
  \label {alg:xoridlshr}
  \KwData{Input file S to share.}
  \KwResult{Three Shares SH1,SH2,SH3 of same size as the original file.}
  \BlankLine
  \While{not at end of the input file}{
  $s$=read\_byte(S)  // read a byte or pixel \\
  $r$=random(257) // random number between 0-256 \\
  $s1,s2$=split\_two($s$)// split s into 2 nibbles \\
  $r1,r2$=split\_two($r$) // split r into 2 nibbles\\
  $s0=0000$ // a dummy variable initialized to zero \\
  $sh1=s0 \oplus r1 || s2 \oplus r2$ 
 // sh1 is the share1 pixel and '$||$' is concatenation operation
  $sh2=s1 \oplus r1 || s0 \oplus r2$
 //sh2 is the share2 pixel and '$||$' is concatenation operation
 $sh3=s2 \oplus r1 || s1 \oplus r2$
  //sh3 is the share3 pixel and '$||$' is concatenation operation
  }
  \caption{(2,3)XOR Ideal Secret Sharing}
  \end{algorithm}
    
\begin{algorithm}[!ht]
  \label {alg:xoridlrec12}
  \KwData{Shares SH1 and SH2}
  \KwResult{Original secret S that is shared}
  \BlankLine 
  \While{not at end of the input files}{
  $sh1$=read\_byte(SH1)  // read a byte or pixel \\
  $sh2$=read\_byte(SH2) \\
  $x1,y1$=split\_two($sh1$) \\
  $x2,y2$=split\_two($sh2$) \\
  $s1=x1 \oplus x2$ \\
  $s2=y1 \oplus y2$ \\
  $s=s1||s2$ 
  }
  \caption{(2,3)XOR Ideal Secret Recovery SH1SH2 }
  \end{algorithm}
  
 \begin{algorithm}[!ht]
    \label {alg:xoridlrec13}
    \KwData{Shares SH1 and SH3}
    \KwResult{Original secret S that is shared}
    \BlankLine 
    \While{not at end of the input files}{
    $sh1$=read\_byte(SH1)  // read a byte or pixel \\
    $sh3$=read\_byte(SH3) \\
    $x1,y1$=split\_two($sh1$) \\
    $x3,y3$=split\_two($sh3$) \\
    $s2=x1 \oplus x3$ \\
    $s1=y1 \oplus y3 \oplus s2$ \\
    $s=s1||s2$ 
      }
    \caption{XOR Ideal Secret Recovery SH1SH3 }
    \end{algorithm}
\newpage  
\begin{algorithm}[!ht]
  \label {alg:xoridlrec23}
   \KwData{Shares SH2 and SH3}
  \KwResult{Original secret S that is shared}
   \BlankLine 
   \While{not at end of the input files}{
   $sh2$=read\_byte(SH2)  // read a byte or pixel \\
   $sh3$=read\_byte(SH3) \\
   $x2,y2$=split\_two($sh2$)\\
   $x3,y3$=split\_two($sh3$) \\
   $s1=y2 \oplus y3$ \\
   $s2=x2 \oplus x3 \oplus s1$ \\
   $s=s1||s2$ 
   }
   \caption{(2,3)XOR Ideal Secret Recovery SH2SH3}
\end{algorithm} 

\section{Conclusion}	
The confidentiality,availability and performance requirement of storage system is addressed in this paper.Secret sharing based solutions provides information theoretic security and also provides trust and reliability.We developed simple XOR based schemes which is easy to implement and increase the performance.The storage requirement can also be reduced if we use scheme where the share size is only half the size of the original secret.The schemes mentioned in this paper are simple and easy to implement when sharing data with third party servers.The cost factor must also be considered.A $(3,2)$ or $(4,2)$  schemes are the best choices.The cost factor can also be reduced by using the non ideal XOR based scheme where the share size is reduced to half but the information theoretic security is compromised.A secret vector which indicates the share number that each server stores can be kept secret.A simple substitution or transposition cipher can also be used as a preprocessing step before sharing the file for additional security .The use of these schemes can be further explored in other areas where the threshold required is as specified in the algorithm. We have used this schemes for efficient sharing of secret images also.
 

\end{document}